\documentclass[aps,twocolumn,superscriptaddress,showpacs,floatfix]{revtex4-1}
\usepackage{graphicx,epsfig,psfrag}
\usepackage{epstopdf}
\usepackage{amsfonts}
\usepackage{times}
\usepackage{graphics,dcolumn,bm,float}
\usepackage{amssymb,amsmath,amsfonts,rotate,color}
\usepackage[colorlinks=true,citecolor=blue]{hyperref}
\usepackage{soul}
\hypersetup{colorlinks=true,citecolor=blue,linkcolor=red,urlcolor=blue}

\begin{document}
\renewcommand{\thefigure}{\arabic{figure}}
\def\clr#1{\textcolor{red}{#1}}
\def\be{\begin{equation}}
\def\ee{\end{equation}}
\def\ber{\begin{eqnarray}}
\def\eer{\end{eqnarray}}
\newcommand{\h}[1]{{\hat {#1}}}
\newcommand{\hdg}[1]{{\hat {#1}^\dagger}}
\newcommand{\bra}[1]{\left\langle{#1}\right|}
\newcommand{\ket}[1]{\left|{#1}\right\rangle}
\newcommand{\la}{\langle}
\newcommand{\ra}{\rangle}
\newcommand{\cd}{c^\dagger}
\newcommand{\nn}{\nonumber}
\newcommand{\RN}[1]{%
\textup{\uppercase\expandafter{\romannumeral#1}}%
}

\title{Impact of topology on the impurity effects in extended $s$-wave superconductors with spin-orbit coupling}

\date{\today}

\author{M. Mashkoori}\email{mahdi.mashkoori@physics.uu.se}
\affiliation{Department of Physics and Astronomy, Uppsala University, Box 530, SE-751 21 Uppsala, Sweden}

\author{A. G. Moghaddam}\email{agorbanz@iasbs.ac.ir}
\affiliation{Department of Physics, Institute for Advanced Studies in Basic Science (IASBS), Zanjan 45137-66731, Iran}
\affiliation{Research Center for Basic Sciences and Modern Technologies (RBST), Institute for Advanced Studies in Basic Science (IASBS), Zanjan 45137-66731, Iran}
\affiliation{School of Physics, Institute for Research in Fundamental Sciences (IPM), Tehran 19395-5531, Iran}

\author{M. H. Hajibabaee}
\affiliation{Department of Physics, Institute for Advanced Studies in Basic Science (IASBS), Zanjan 45137-66731, Iran}

\author{A. M. Black-Schaffer}\email{annica.black-schaffer@physics.uu.se}
\affiliation{Department of Physics and Astronomy, Uppsala University, Box 530, SE-751 21 Uppsala, Sweden}

\author{F. Parhizgar}
\affiliation{Department of Physics and Astronomy, Uppsala University, Box 530, SE-751 21 Uppsala, Sweden}
\affiliation{School of Physics, Institute for Research in Fundamental Sciences (IPM), Tehran 19395-5531, Iran}

\begin{abstract}
We investigate the impact of topology on the existence of impurity subgap states in a time-reversal-invariant superconductor with an extended $s$-wave pairing and strong spin-orbit coupling. By simply tuning the chemical potential we access three distinct phases: topologically trivial $s$-wave, topologically non-trivial $s_\pm$-wave, and nodal superconducting phase. 
For a single potential impurity we find subgap impurity bound states in the topological phase, but notably no subgap states in the trivial phase. This is in sharp contrast with the expectation that there would be no subgap state in the presence of potential impurities in $s$-wave superconductors.
These subgap impurity states have always finite energies for any strength of the potential scattering and subsequently, the superconducting gap in the topological $s_\pm$-wave phase survives but is attenuated in the presence of finite disorder. By creating islands of potential impurities we smoothly connect the single impurity results to topological edge states of impurity island.
On the other hand, magnetic impurities lead to the formation of Yu-Shiba-Rusinov states in both the trivial and topological phases, which even reach zero energy at certain scattering strengths. 
We thus propose that potential impurities can be a very valuable tool to detect time-reversal-invariant topological superconductivity.

\end{abstract}

\maketitle
\section{Introduction}

Questions about the role of impurities for the properties of materials have for a long time been one of the most important challenges in condensed matter physics \cite{ziman,anderson58,lee85}. This is also true for superconductors (SCs), where  many different features have already been attributed to impurity effects \cite{balatsky}. Already soon after the development of BCS theory, conventional spin-singlet $s$-wave superconductivity was found to be very robust against the random potential disorder induced by non-magnetic impurities, a result coined as \emph{Anderson's Theorem} \cite{anderson59,anderson59-2,abrikosov59,skalski}. In contrast, in the presence of time-reversal symmetry breaking perturbations, such as magnetic impurities, or when the superconducting gap has a nodal structure, like $p$-wave or $d$-wave pairing symmetry, impurity scatterings has been found to suppress superconductivity \cite{gorkov61, kadanoff, hirschfeld}. For example, the competitive interplay of magnetism and superconducting orders in the vicinity of a single classical spin impurity, has been shown to give rise to so-called subgap Yu-Shiba-Rusinov (YSR) states \cite{yu,shiba,rusinov}. These spin-polarized bound states have been experimentally extensively verified using low-temperature scanning tunneling microscopy (STM) and scanning tunneling spectroscopy (STS) techniques \cite{yazdani97, ji08, menard15, ruby16}.
\par  
Recently, the interest in impurity effects in SCs has been further boosted by the theoretical predictions and subsequent experiments suggesting that topological superconductivity can emerge in an atomic chain of magnetic impurities on top of a conventional SC \cite{choy11, nadj-perge13, nadj-perge14,nagaosa13, simon13,loss13,glazman13,das-sarma14, macdonald14}. 
Topological SCs are particularly interesting as they host exotic excitations such as Majorana bound states with promising applications in fault-tolerant quantum computations \cite{kitaev01,fu08,nayak08,qi-zhang,alicea12,beenakker13}. The most well known example of topological superconducting phases is the spinless chiral $p$-wave phase, in which time-reversal symmetry is broken and this can be engineered in semiconductor/SC heterostructures \cite{das-sarma10,oreg10,kouwenhoven12,kouwenhoven16} or with magnetic impurities as mentioned above. But, based on the periodic table of topological phases many other possibilities exists\cite{schneyder08}. In particular, topological \emph{time-reversal-invariant} (TRI) superconductivity in class DIII has recently been intensively studied, where Kramer's pairs of Majorana bound states are predicted to exist \cite{schneyder08,kitaev09,schneyder16,Klinovaja2014-1,Klinovaja2014-2,Klinovaja2014PRL,Arrachea2017,Arrachea2018}.
Beside many proposals for TRI topological SCs exploiting exotic interactions or complex structures \cite{raghu09,hughes10,berg10,ando11, nakosai12,berg13,haim14,law14,schmalian15}, a more simple route has been proposed based on proximity effect between Rashba semiconductors and $s_\pm$-wave SCs. \cite{zhang13}. Unconventional $s_\pm$ pairing is believed to exist in the iron-based SCs and is characterized by two separate $s$-wave gaps with opposite signs in different parts of the first Brillouin zone \cite{mazin08,greene10,mazin10,mazin11,
stewart11}.
\par
Motivated by both theoretical predictions and experimental observations of topological superconductivity, various works have explored the effects of both magnetic and non-magnetic impurities on the order parameter and bound states formation in chiral $p+ip$-wave and other time-reversal symmetry broken phases \cite{Sau2013,Slager2015,Hyart2015,Hyart2016,KaladzhyanPRB,KaladzhyanJPCM,Mashkoori2017,Cayao2015}. In addition, general symmetry analyses of the existence of zero-energy impurity states have been performed in both topological insulators \cite{Slager2015} and superconductors \cite{Hyart2015,Hyart2016}.
Furthermore, impurity-related phenomena in iron-based SCs have also been the subject of intense investigation on on theory and experiment fronts \cite{Matsumoto2009,Bang2009,Hirschfeld09,Senga09,Akbari2010,Gastiasoro2013,Yin2015, Jiao2017}. Specifically, exploiting a multi-band model, it has been shown that it is needed to invoke interband scattering
in $s_\pm$-wave SCs to generate potential impurity bound states \cite{Matsumoto2009}.
\par 
In this work, we turn the attention to impurity effects in a two-dimensional Rashba spin-orbit semiconductor together with an extended $s$-wave pairing. This system has multiple advantages. It can easily be realized in hybrid structures combining a Rashba spin-orbit coupled semiconductor with an $s_\pm$-wave iron-based SC \cite{zhang13}. It also has a wide range of tunability, where it is very simple to tune from a topological non-trivial to a trivial regime, and also to incorporate a nodal phase in-between. This offers exemplary opportunities for both theory and experiment to investigate the impact of topology on impurity states, where the superconducting state itself has the normally expected very robust $s$-wave symmetry.  
\par
As a key finding we reveal distinct behaviors of potential impurities in topologically trivial and non-trivial phases of the system. In fact, while a non-magnetic impurity does not induce any subgap states in the trivial phase, subgap bound states emerges in the topological phase.  Nevertheless, non-magnetic impurities never give rise to zero-energy bound states and they thus only moderately suppress pairing in topological phase of the system, but never fully deplete it. 
This is in sharp contrast to the behavior of standard $s$-wave SCs for which no subgap state is induced as a result of interaction with a non-magnetic impurity. 
We are able to attribute these subgap states to the bulk-boundary correspondence of topological matter \cite{Sato2017}, viewing the impurity region as a trivial domain within the topological non-trivial bulk. The connection becomes even more evident when studying finite size islands of potential impurities.
Moreover, we show that, irrespective of topology, adding a magnetic impurity results in YSR subgap states.
Our results both demonstrate that potential impurities can be used as a probe of the topology in TRI Rashba coupled SCs and establish that interband scattering is not the only mechanism in which potential impurities generate sub gap states in $s_\pm$-wave SCs as stated in Ref.~\cite{Matsumoto2009}.
\par
The remainder of the paper is organized as follows. In Sec.~\ref{sec-model}, we introduce the system and its Hamiltonian and discuss its distinct phases, as well as describe both the $T$-matrix and numerical lattice diagonalization formalisms used to study the impurity effects. In Sec.~\ref{sec-result} we first present the results for single magnetic and potential impurities. Then, we discuss small island of impurities. Finally, the summary and concluding remarks are presented in Sec. \ref{sec-conc}.
\section{Model and method}\label{sec-model}
In this section, we first introduce the model Hamiltonian of the system under study, namely, a two-dimensional Rashba spin-orbit coupled material with an extended $s$-wave SC pairing and then present our methods for solving the impurity scattering problem. This system can be realized in a hybrid structure composed of a thin Rashba layer in proximity to an $s_\pm$-wave superconductor. A large tunability of this hybrid structure take it through three distinct phases: TRI topological $s_\pm$-wave, nodal, and trivial $s$-wave superconducting phase and makes it a very promising model to study topological effects. To in-detail investigate impurity scattering, we implement two different methods: the $T$-matrix approach, which is performed in momentum space, and tight-binding (TB) calculations in the real-space lattice basis for a finite size square lattice.
\subsection{Bulk Hamiltonian}
A TRI topological SC can be constructed at the interface of a two-dimensional (2D) Rashba material with an $s_{\pm}$-wave spin singlet SC, as originally proposed in Ref.~\cite{zhang13}. In order to model such an interface, we assume a 2D square lattice with nearest neighbor hoping $t$, Rashba spin-orbit interaction $\lambda_R$, and superconducting pairing which consists of on-site $\Delta_0$ and isotropic nearest neighbor $\Delta_1$ terms, which give extended $s$-wave symmetry.  
The TB Hamiltonian within the standard mean-field framework for superconductivity reads, 
\begin{align}
{\cal H}_0^{TB}=
& 
\sum_{ {\bf ij} \sigma}
[ -\frac{1}{2}(t_{{\bf ij}} + \mu\delta_{{\bf ij}}) c_{{\bf i}\sigma}^\dag
c_{{\bf j}\sigma}+{\Delta_{{\bf i}{\bf j}}}c_{{\bf i}\sigma}^\dag
c_{{\bf j} {\bar\sigma}}^\dag ]\nonumber \\ 
& -\lambda_R \sum_{\bf i ,\eta=\pm} \eta c_{{\bf i},\uparrow}^{\dag} 
(c_{{\bf i}-\eta\hat{\bf x},\downarrow}-
ic_{{\bf i}-\eta\hat{\bf y},\downarrow}) + \textrm{H.c.} ,
\label{tight-binding.eq}
\end{align}
with $\Delta_{{\bf i}{\bf j}}=\Delta_0$ for ${\bf i}={\bf j}$, while we restrict the range of hopping and pairing to nearest neighbors as $t_{\bf{i}\neq {\bf j}} = t$ and  $\Delta_{{\bf i}\neq{\bf j}}=\Delta_1$. 
Since superconductivity is induced by proximity effect, e.g.~ from an iron-based SC substrate, self-consistency should not play a major role, even for the electronic structure of impurities (see e.g.~Ref.~\cite{Lothman2014, Mashkoori2017, Awoga2018}), and we can safely assume constant order parameters.

The resulting Bogoliubov-de Gennes (BdG) Hamiltonian in momentum space for the fully translationally invariant bulk is,
\ber
&&{\cal H}_0 = \sum_{\bf k} \Psi_{\bf k}^\dag h_{\bf k} \Psi_{\bf k},\nonumber\\
&&h_{\bf k} = \tau_3\left[ \sigma_0{\xi_{\bf k} + 2\lambda_R \left( \sigma_x \sin k_y - \sigma _y
\sin k_x \right)} \right] + \tau_1\sigma_0\Delta_{\bf k},~~~
\label{bdg}
\eer
where we define
\ber
&&\label{xi_k}\xi_{\bf k} =  - 2t (\cos k_x + \cos k_y)  - \mu, \nonumber\\
&&\label{Delta_k}\Delta_{\bf k} = {\Delta _0} + 2{\Delta _1}\left( {\cos {k_x} + \cos {k_y}} \right).
\eer
Here, $\sigma_i$ and $\tau_i$ are the Pauli matrices in spin and Nambu (particle-hole) spaces, respectively.  The so-called Nambu spinor $\Psi_{\bf k}^T = ( 
c_{{\bf k} \uparrow },c_{{\bf k} \downarrow },c_{ - {\bf k} \downarrow }^\dag , - c_{-{\bf k} \uparrow }^\dag  )$, consists of four elements corresponding to annihilation and creation of electrons with different spins. 
The extended $s$-wave pairing $\Delta_{\bf k}$ in Eq.~\eqref{Delta_k} can lead to a 
sign-changing order parameter between the $\Gamma(0,0)$ and the $M(\pi ,\pi)$ points in the Brillouin zone, which subsequently leads to the topologically non-trivial $s_\pm$-wave superconducting phase. The significance of this is that it provides a very rich phase diagram including TRI topologically trivial, non-trivial, and nodal phases but still within a single spin-full band picture. It is worth mentioning that according to the presence of strong spin-orbit coupling in the system, all these phases are achievable for a wide range of parameters. 
\par
\begin{figure}[t]
\centering
\includegraphics[width=0.23\textwidth]{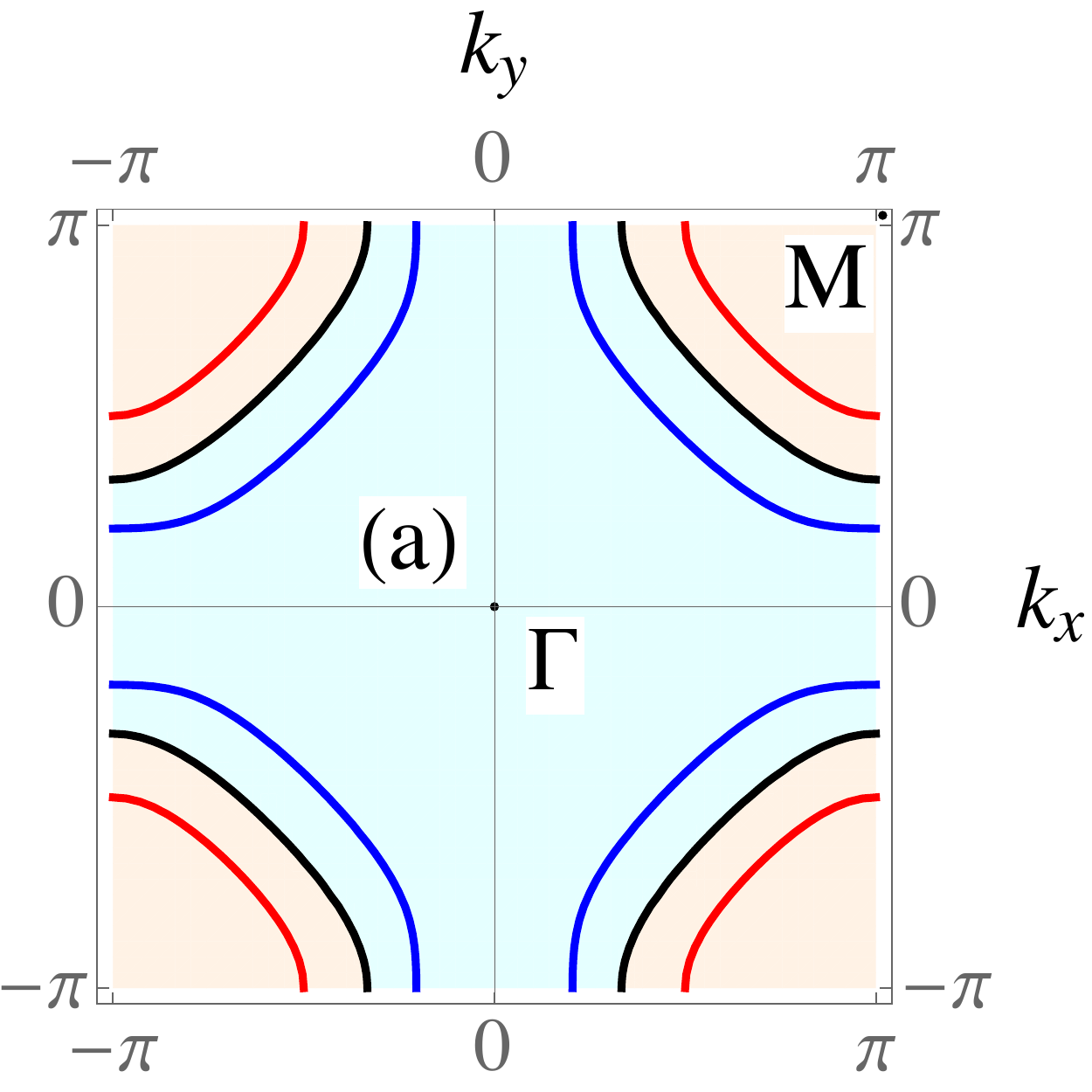}
\includegraphics[width=0.23\textwidth]{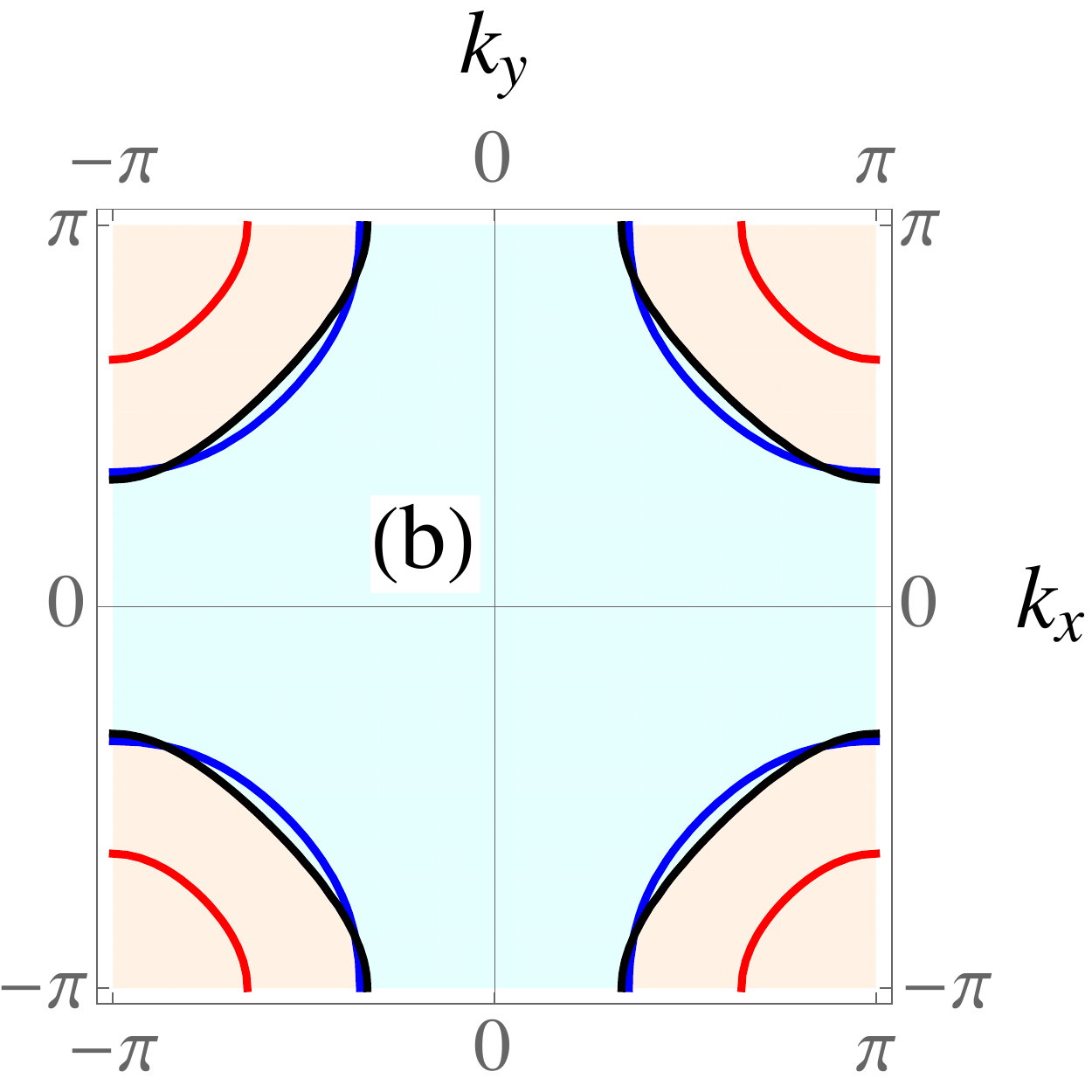}
\includegraphics[width=0.3\textwidth]{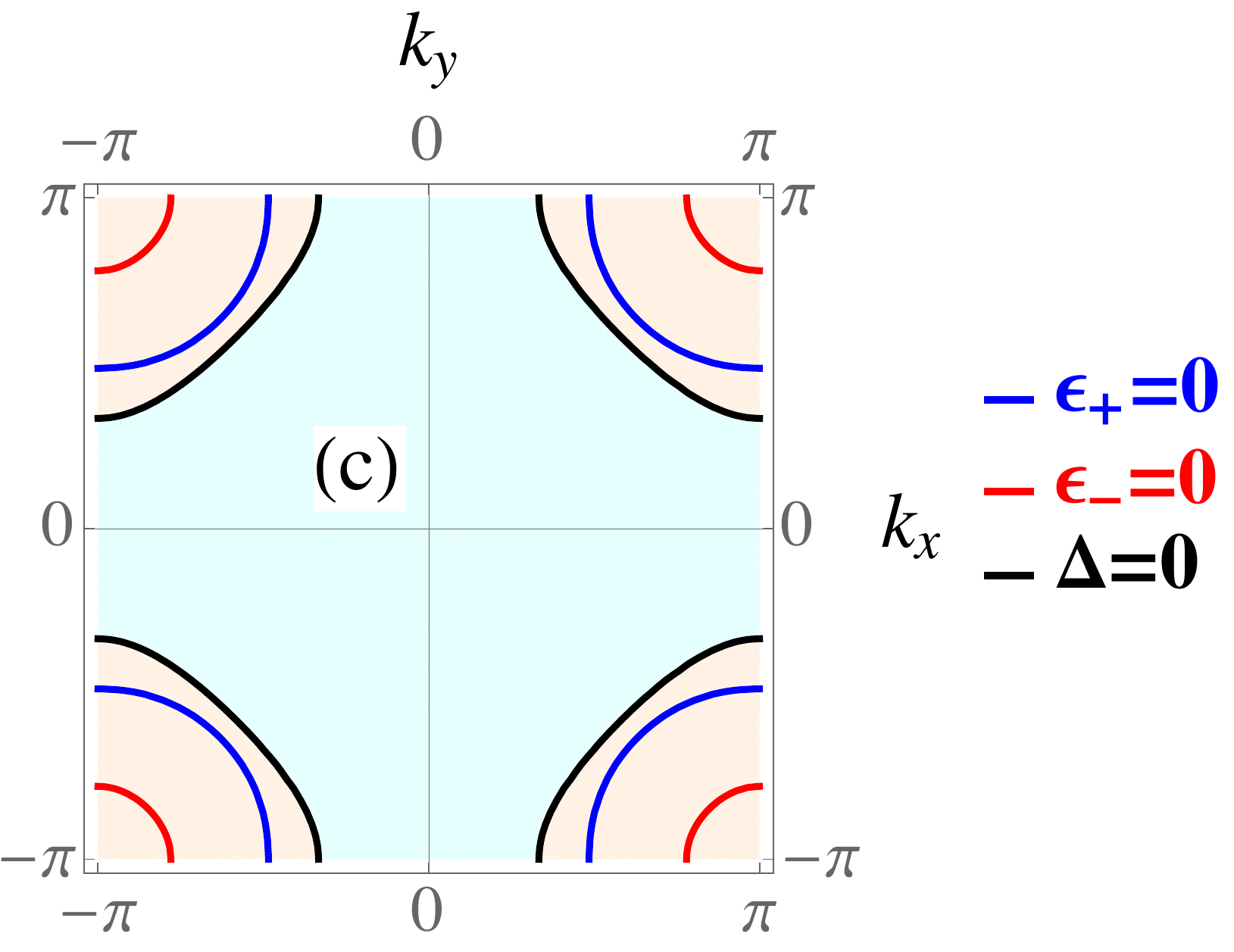}
\caption{(Color online.) Fermi surfaces $\varepsilon_\pm({\bf k})=0$ and nodal-lines of the gap function $\Delta_{\bf k}$ in the topologically non-trivial (a), nodal (b), and trivial (c) phases.}
\label{FS.fig}
\end{figure}

The energy dispersion of the Hamiltonian Eq.~\eqref{bdg} reads
\ber
&&E_\pm ({\bf k}) =  \pm \sqrt{\varepsilon_\pm^2({\bf k})   + |\Delta_{\bf k}|^2}, \label{ekbdg}\nonumber\\
&&\varepsilon_\pm({\bf k})=
\xi_{\bf k}  \pm 2\lambda_R\sqrt{\sin^2 k_x + \sin ^2 k_y}.\label{varep}
\eer
The Fermi surface consists of two portions, which encircle either the $\Gamma$ or $M$ points and are given by the solutions of the equation $\varepsilon_\pm({\bf k})=0$. The superconducting gap function also reveals a nodal curve governed by $\Delta_{\bf k}=0$, along which the superconducting gap confronts a sign-change. Fig.~\ref{FS.fig} shows the Fermi surface of the normal Hamiltonian $(\varepsilon_\pm=0)$ together with nodal lines of the gap function $(\Delta_k=0)$. 
\par
As shown in Fig.~\ref{FS.fig}(a,c), two different gapped phases can be obtained, depending on whether both Fermi lines lie completely inside a region with a single sign of the pairing potential $\Delta_{\bf k}$ (in Fig.~\ref{FS.fig}(c)) or if they lie in different regions with different signs of $\Delta_{\bf k}$ (Fig.~\ref{FS.fig}(a)). The former is a topological trivial conventional $s$-wave SC where the order parameter does not change sign, while the latter corresponds to a topological $s_{\pm}$-wave SC. The topology can be identified by observing that the system belongs to class {D\RN{3}} of topological states \cite{schneyder08} which has a $\mathbb{Z}_2$ topological index $\nu=0 (1)$ for the topologically trivial (non-trivial) phases, respectively, and is protected by time-reversal, particle-hole, and chiral symmetries. 
A gapless (nodal) phase also takes place when one of the Fermi pockets crosses this nodal curve, as illustrated in Fig.~\ref{FS.fig}(b). It is clear that in this phase, there exist multiple points in the Brillouin zone with $E_\pm({\bf k})=0$ which gives rise to gapless or nodal superconducting state. 
The large benefit of our model is thus that three distinct phases are easily achievable by simply tuning the chemical potential which moves the Fermi surface positions. 
\par
\begin{figure}[t]
\centering
\includegraphics[width=0.5\textwidth]{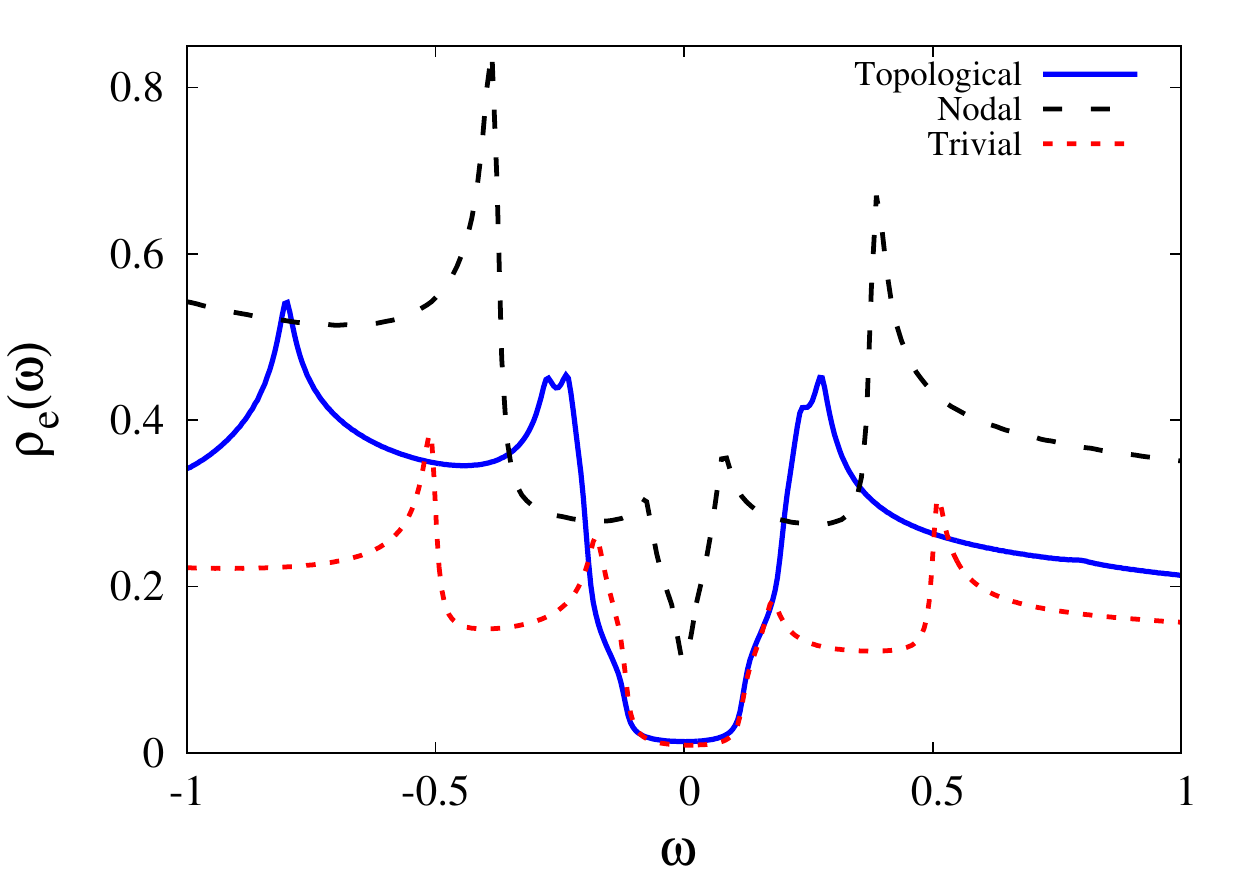}
\caption{(Color online.)
Electronic DOS in the bulk for Eq.~\eqref{bdg}, in the topologically non-trivial (blue solid), nodal (black dashed), and trivial (red dotted) phases.}
\label{eDOS.fig}
\end{figure}

To obtain the spectral properties for the system, we utilize the bulk Green's function which can be calculated from the Hamiltonian Eq.~\eqref{bdg} as
\begin{align}
G_0({\bf k},\omega) &=\frac{1}{2} \sum_{\eta=\pm} 
\frac{ \omega \tau_0 + \varepsilon_\eta( {\bf k})\tau_3 + \Delta({\bf k})\tau _1
}
{ \omega ^2 - \varepsilon_\eta^2 ( {\bf k}) - |\Delta ({\bf k})|^2} \nonumber \\
& \otimes (
\sigma_0 +\eta \sigma _1 \sin \phi_{\bf k}  -\eta \sigma_2 \cos \phi_{\bf k} ),
\label{G0}
\end{align}
where $\phi_{\bf k}=\arcsin\left( \sin {k_y}/ \sqrt{\sin^2 k_x + \sin^2 k_y}\right)$. Fig.~\ref{eDOS.fig} shows the electronic density of states (DOS) in the bulk calculated from
\begin{equation}
\rho_e(\omega)=\frac{-1}{\pi}\sum_{\bf k} {\rm Im}\left\{ {\rm Tr}\, [(\frac{\tau_0+\tau_z}{2})G({\bf k},\omega)] \right\}.
\end{equation}
and in all three phases; fully gapped topological non-trivial $s_\pm$- and trivial $s$-wave states, as well as in the nodal phase.
Throughout this work, we set the hopping parameter $t = 1$, order parameters $\Delta_0=\Delta_1=0.2$ and Rashba spin-orbit coupling $\lambda = 0.5$. Moreover, we choose $\mu =1,2$, and $2.9$ to represent the topologically non-trivial, the nodal, and the topologically trivial phases respectively, unless we clearly mention other choices of parameters. 
This choice of parameters is both convenient and, most importantly, generates the same size gaps in both the trivial and non-trivial topological phases. The nodal phase appears in between and illustrates the topological phase transition. As seen in Fig.~\ref{eDOS.fig}, it is thus not possible to distinguish between the two gapped phases by only studying the DOS, but other means are needed.
We note that several other choices of parameters can be used to model the three different phases, but our conclusions  remain intact.
 
\begin{figure}[htb]
\centering
\includegraphics[width=0.37\textwidth]{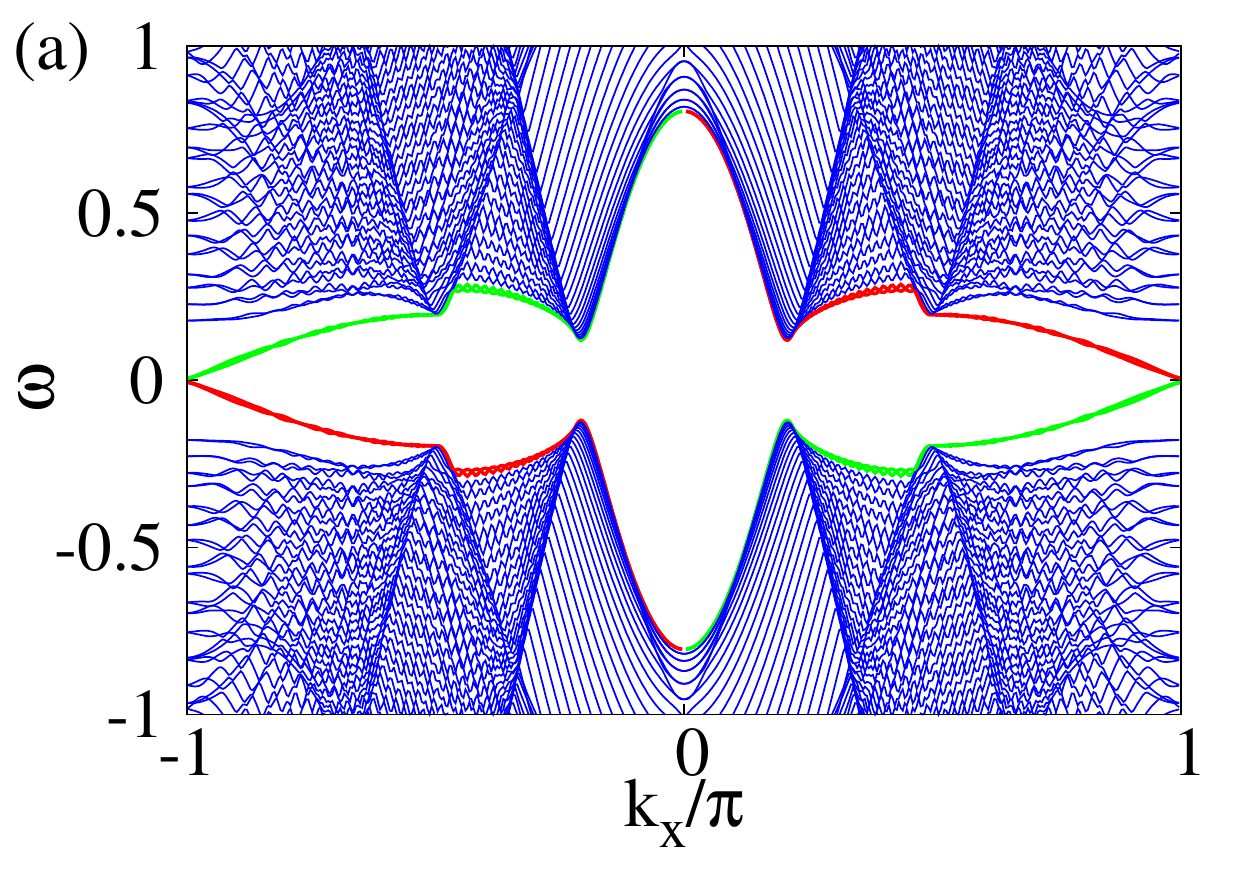}
\includegraphics[width=0.37\textwidth]{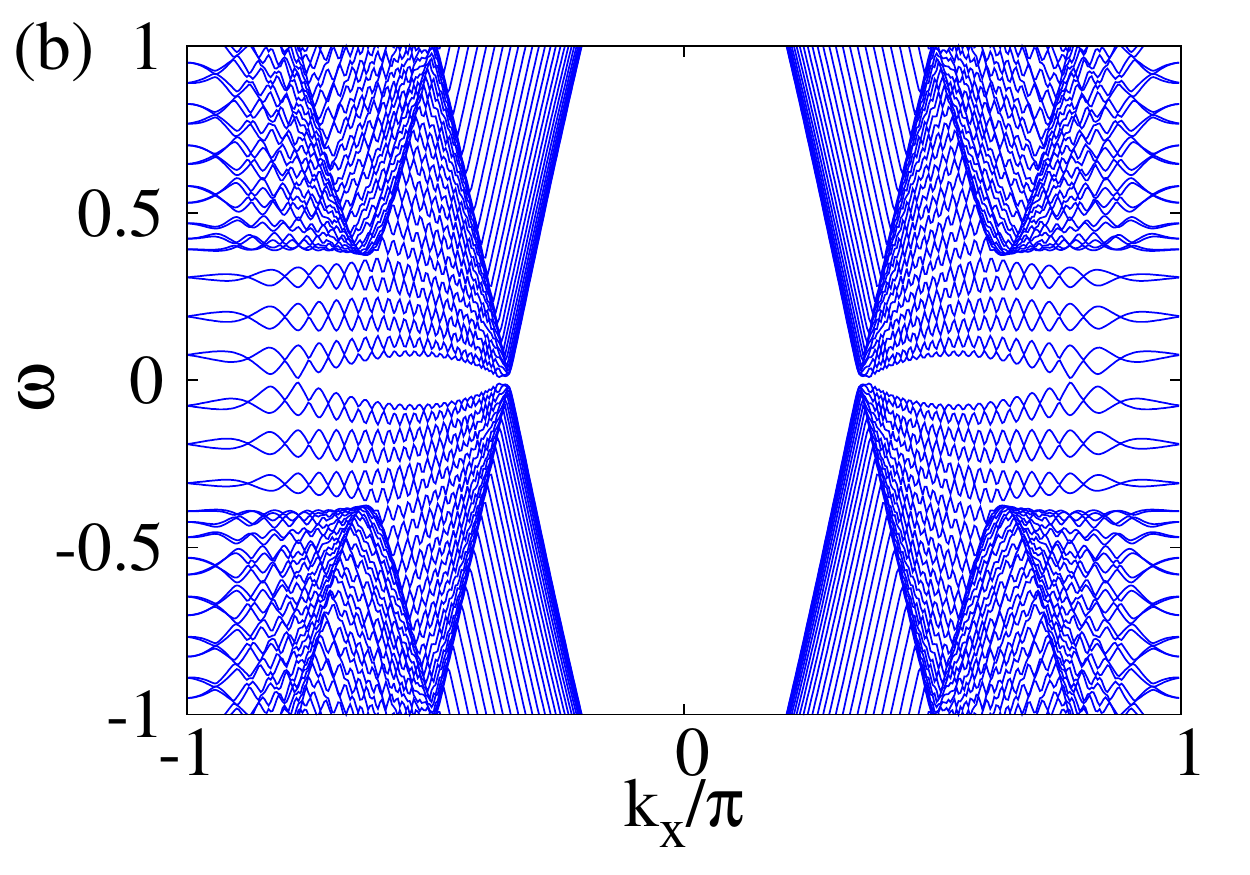}
\includegraphics[width=0.37\textwidth]{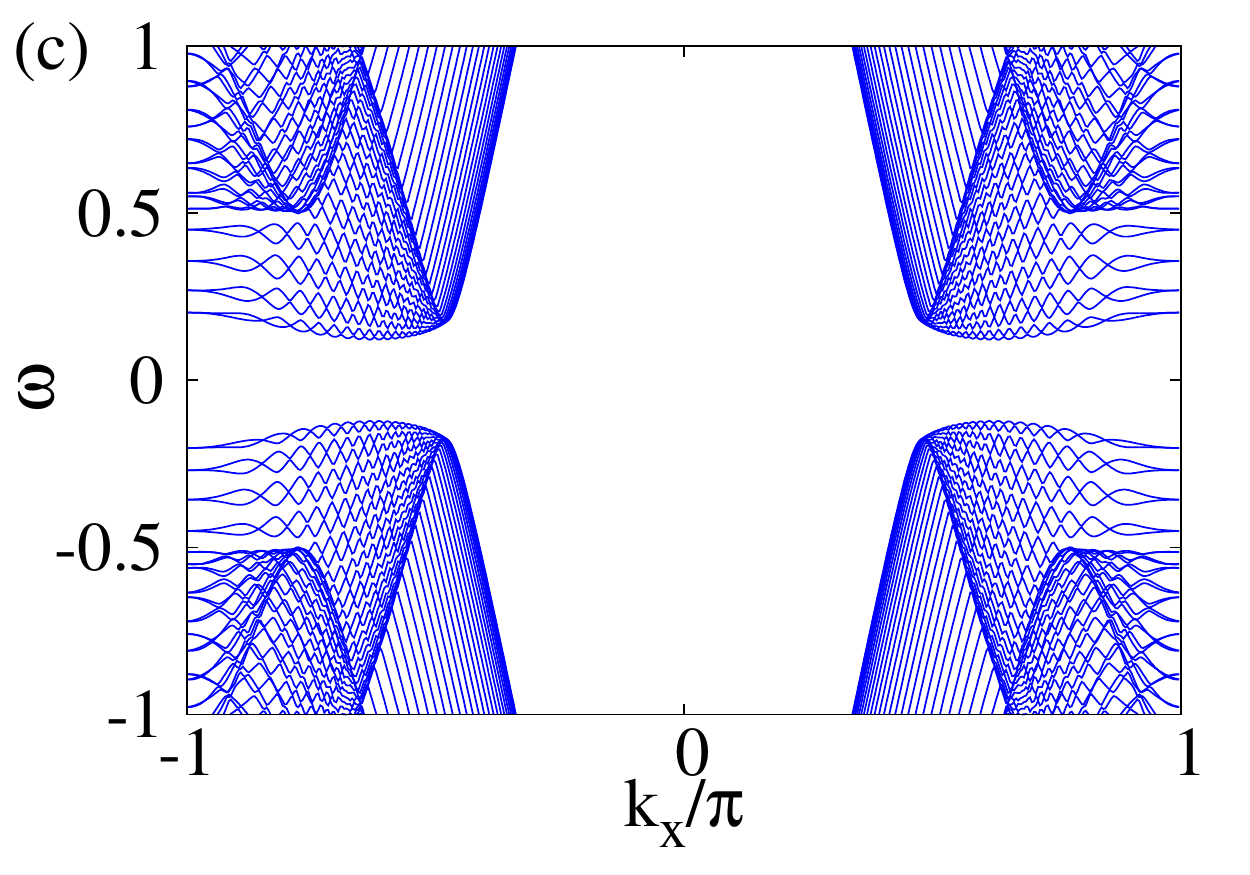}
\caption {(Color online.) Energy dispersion of a 50 lattice site wide nano-ribbon of Eq.~\eqref{tight-binding.eq} in the topologically non-trivial (a), nodal (b), and topologically trivial (c) phases. A Majorana Kramers pair appear in the (a) as signature of
non-trivial topology, with left (red) and right (green) moving Majorana states. }
\label{Disp.fig}
\end{figure}

In order to capture the edge modes in the different phases, we instead find the spectrum of the real space Hamiltonian Eq.~\eqref{tight-binding.eq} for a 50 lattice sites wide nano-ribbon, as shown in Fig.~\ref{Disp.fig}.
While the nodal phase is clearly gapless also for a nano-ribbon as we see in Fig.~\ref{Disp.fig}(b), the topologically non-trivial, Fig.~\ref{Disp.fig}(a), and trivial, Fig.~\ref{Disp.fig}(c), phases both have a band gap of $2\Delta_g \simeq 0.25$.
In the topological phase helical edge states also appear at the nano-ribbon edges. These are illustrated in Fig.~\ref{Disp.fig}(a) by red (green) color, indicating right (left) moving states. Due to time-reversal symmetry these edge states emerge as Majorana Kramers pairs. 

\subsection{Impurity scattering}
We add a single impurity into the system and compare the results in different topological phases of the system. For magnetic impurities we treat the spin of the impurity as a classical vector $\hat{\bf S}$ and neglect quantum fluctuations. Thus the impurity can be modeled as
\begin{equation}
{\cal H}_{\rm imp} = \sum_{\sigma\sigma'}
c_{{\bf a} \sigma}^{\dag}   \left(U\delta_{\sigma\sigma'} - {J}\hat{\bf S}\cdot {\bm \sigma}_{\sigma\sigma'} \right)  c_{{\bf a} \sigma'},
\label{sd.eq}
\end{equation}
where $U$ is the strength of the impurity potential, $J$ the magnetic coupling, and ${\bf a}$ indicates the position of the impurity in the lattice. We find that the energy of any bound states as well as their emergence do not depend on the impurity spin orientation since we only consider the first harmonic in the scattering \cite{Kim2015}.
Thus, we assume the impurity moment to be aligned in $z$-direction and the Hamiltonian can be written as
${\cal H}={\cal H}_0+{\cal H}_{\rm imp}$, where
\begin{align}\label{eq:himp}
{\cal H}_{\rm imp}&=\sum_{\bf k k'} \Psi^\dag_{\bf k} 
\hat h_{\rm imp}
   \Psi_{\bf k'}, \nonumber
\\
\hat h_{\rm imp}&
=U\tau_3\sigma_0 + J \tau_0\sigma_3.
\end{align}
The system in Eq.~\eqref{eq:himp} and more generally any location dependent perturbation can be treated using the $T$-matrix formalism. The Green's function defined as $ {\bf G}=(\omega-{\cal H}_0-{\cal H}_{\rm imp})^{-1}$
is then expanded using the Dyson equation to we obtain the Green's function in momentum representation as
 below,
\begin{align}
G({\bf k},{\bf k}',\omega)&=G_0({\bf k},\omega)\delta_{{\bf k},{\bf k}'}
+G_0({\bf k},\omega)T(\omega)G_0({\bf k}',\omega),\nonumber \\ \label{tmatrix-1}
T(\omega)&=[{1}-h_{\rm imp}\sum_{\bf k} G_0({\bf k},\omega)]^{-1}h_{\rm imp},
\end{align}
where $G_0({\bf k},\omega)= (\omega - h_{\bf k})^{-1}$ is the bare Green's function given in Eq.~\eqref{G0}. Notice that singularities of the $T$-matrix immediately gives the impurity induced bound states.

We also perform numerical lattice calculations and we directly solve Eq.~\eqref{tight-binding.eq} on a finite lattice with $L_x = L_y=50$ lattice points employing periodic boundary conditions to avoid possible hybridization of subgap states with Majorana edge states in the topological phase. We have checked that this supercell size is large enough to prevent any hybridization between impurities in different supercells. The maximum size of the impurity islands are chosen to be 25 impurities in a $5 \times 5$ square and placed to preserve rotation symmetry of the system. The full system including impurities is then solved by exact diagonalization.

\section{Results and discussion}\label{sec-result}
Having presented the system and established its different topological phases in the bulk in the previous section, we now turn to the main purpose of this work and study the effects of impurities and how they can be used to discern topology.
To tackle the problem of a single impurity, we first use the momentum space representation of the $T$-matrix formalism given by Eq.~(\ref{tmatrix-1}) to find the local density of states (LDOS) close to the impurity. Then we exploit the lattice model based on the real-space Hamiltonian in Eq.~\eqref{tight-binding.eq} to gain complementary information. We finally turn to investigating islands of impurities to connect the single impurity limit to properties of edge states.
Before proceeding we note that the two approaches give the same result for the same system in the thermodynamic limit, but, the numerical investigation of the lattice model is more convenient for certain studies and especially for impurity islands.

\subsection{Momentum-space $T$-matrix approach}
\begin{figure}[t]
\centering
\includegraphics[width=0.95\linewidth]{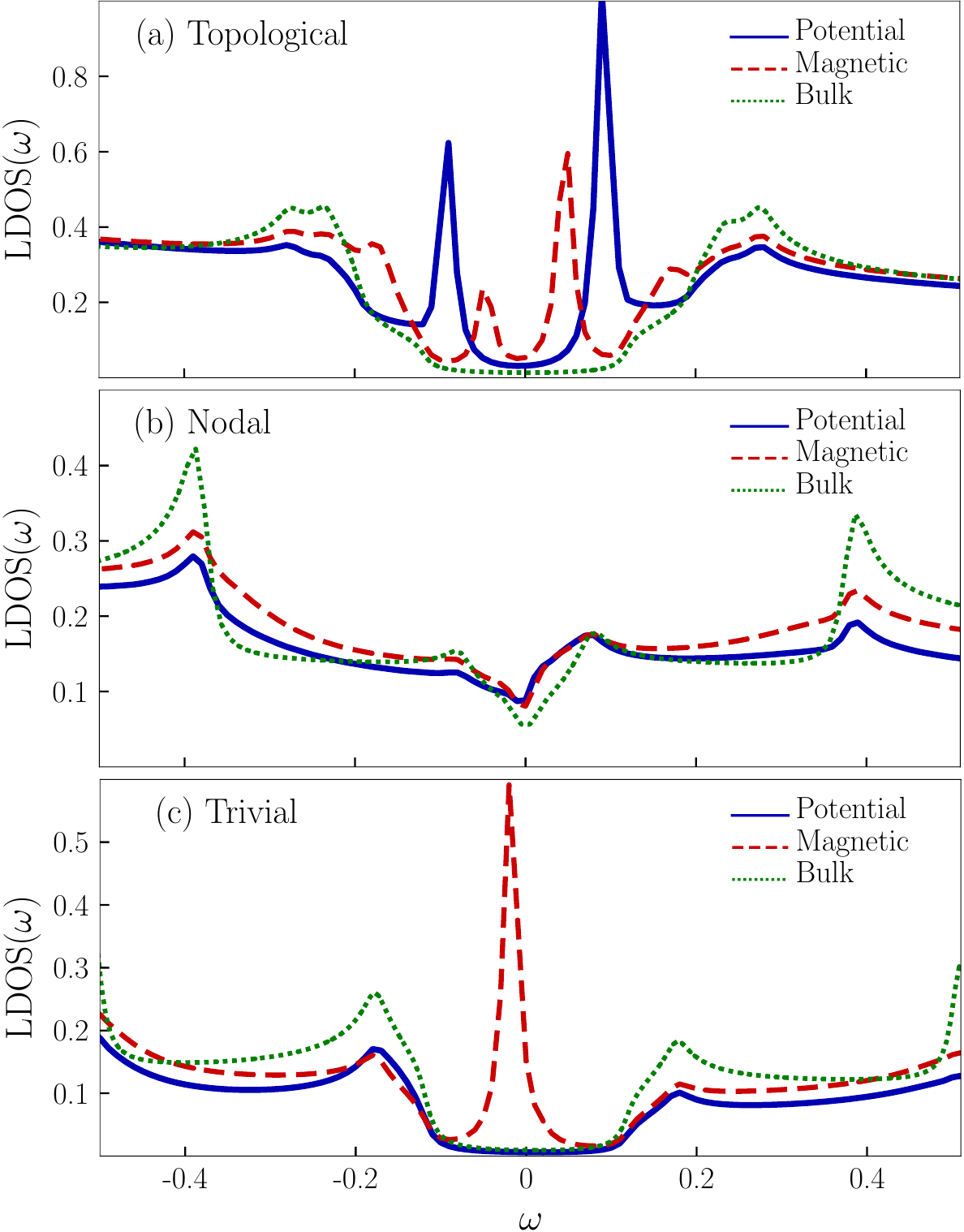}
\caption{(Color online.) LDOS at the impurity site for potential ($U =3$) and magnetic impurities ($J = 3$) in the three different phases, topologically non-trivial (a), nodal (b), and topologically trivial (c).}
 \label{M5.fig}
  \end{figure}

We start our survey by studying the LDOS at the vicinity of a single magnetic and non-magnetic impurity embedded in the 2D Rashba semiconductor with extended $s$-wave pairing. Using the impurity Hamiltonian $h_{\rm imp}$, introduced in Eq.~\eqref{eq:himp} for the magnetic and potential scattering, we obtain the real space Green's function and LDOS, subsequently from the T-matrix formulation in Eq.~\eqref{tmatrix-1}. 
In Fig.~\ref{M5.fig} we show the resulting LDOS in all three phases in the bulk (dotted green) and for pure magnetic (dashed red) and potential (solid blue) scattering, when we set the magnetic (potential) strength to $J=3$ ($U=3$).
Although both topologically trivial and nontrivial phases of the system have similar gapped electronic DOS, as can be seen in Fig.~\ref{eDOS.fig}, the impurity scattering is dramatically different. 

A magnetic impurity induces states inside the superconducting gap in both the topological $(\nu = 1)$ and trivial $(\nu = 0)$ phases. These states are the well-known YSR states, which are formed due to the local breaking of time-reversal symmetry (TRS). Much more interestingly is, however, that a pure potential impurity, which preserves time-reversal symmetry of the system, can generate subgap bound states in the topological phase, as clearly seen in Fig.~\ref{M5.fig}(a). 
These subgap states are clearly different from YSR states, which appear only in the presence of a classical spin (or equivalently tiny localized magnet) and upon TRS breaking. While the potential-impurity induced subgap states are two-fold degenerate, YSR states are non-degenerate and spin-polarized. The existence of these subgap states is  also completely contrary to the usual expectation that there can be no state inside the gap for $s$-wave SCs. 
It is here worth mentioning that subgap impurity states has previously been found to appear in multiband $s_{\pm}$-wave SCs in the presence of non-magnetic impurities \cite{Matsumoto2009,Bang2009,Hirschfeld09,Senga09}. However, the appearance of subgap states in these mutliband SCs is due to interband scattering processes induced by the potential impurities.
In contrast, in our topological $s_\pm$-wave SC, no scattering takes place between the two helical Fermi surfaces and there is also no interband impurity scattering, as the topological features protected by TRS prohibits any such interband scattering. In fact, one can straightforwardly show that any superficially enforced real-valued interband scattering results in a TRS breaking term, equivalent to scattering from a magnetic impurity. Thus, per definition, potential impurities in helical TRI SCs as studied here cannot give rise to interband processes and the sub gap states we find are thus not of the same origin as those previously found in multiband $s_{\pm}$-wave SCs.

In addition, in the trivial phase shown in Fig. \ref{M5.fig}(c), only magnetic impurities induce subgap states and the gap is fully protected against potential scattering in the same way as any conventional $s$-wave SC.
Therefore, we deduce that the appearance of the bound states induced by potential impurity scatterings is a signature of the topological features of the host superconducting material. It should also be mentioned that the topologically non-trivial phase is a result of strong spin-orbit coupling. In the $\lambda_R=0$ limit corresponding to a topologically trivial phase the LDOS instead resembles that of the trivial phase in Fig.~\ref{M5.fig}(c). This further supports that potential impurities can be used to distinguish between different topological classes in spin-orbit coupled extended $s$-wave SCs. 
Possibly related, potential impurities has been shown to generate subgap states in topological $d+id'$-wave SCs with a small $d'$ parameter, while no subgap states are present in the trivial $d+is$ state, when there is instead a subdominate $s$-wave part \cite{Mashkoori2017}.  
Contrary to the gapped phases, in the nodal phase the presence of any impurity does not give rise to any strong fingerprints and LDOS close to the impurity is only changed by small amounts close to the Fermi energy, as seen in Fig.~\ref{M5.fig}(b). 

\subsection{Numerical results of TB model}
In the following we apply complementary numerical lattice calculations. 
\subsubsection{Single impurity}
\begin{figure}[t]
\centering
\includegraphics[width=.65\textwidth]{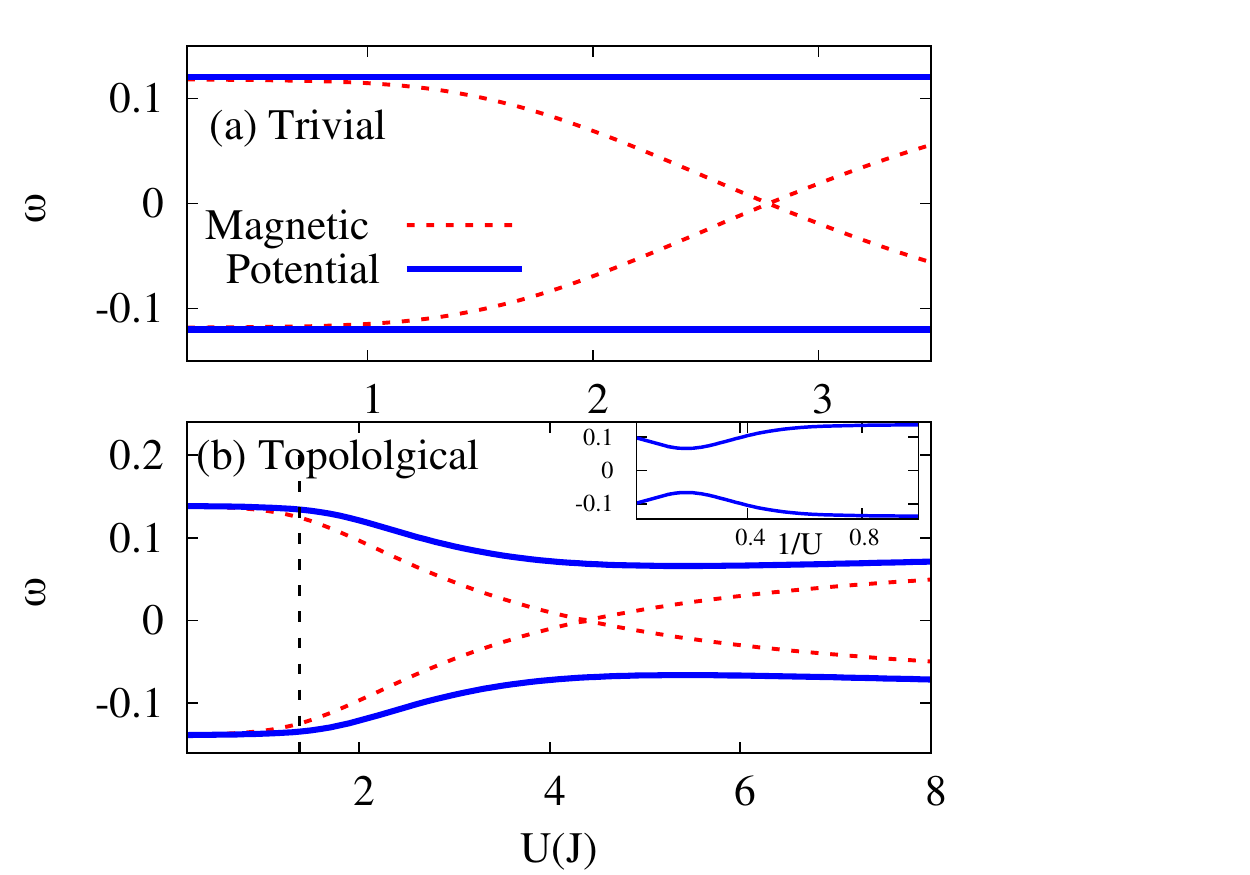}
\caption
{Energy of subgap states from magnetic (red dashed) and potential (blue) impurities in  the topologically trivial (a) and non-trivial (b) phases as function of impurity scattering strengths. Inset of panel (b) shows subgap states as a function of $1/U$.}
 \label{M1.fig}
\end{figure}

As expected from our T-matrix discussion, we find in the trivial phase that  a magnetic impurity leads to the formation of a pair of YSR-states, while scattering from a potential impurity never induce any subgap state.
In Fig.~\ref{M1.fig}(a) we show how the YSR-states from a magnetic impurity crosses zero energy for a critical coupling $J_c$ and then tend towards the gap edge as we increase $J$. For the case of magnetic impurities, the subgap states behave qualitatively similar in the topological and trivial phases. 
For the case of potential impurity, no subgap states emerge for any coupling strength $U$ in the trivial $s$-wave phase.
In contrast, in the topological $s_\pm$-wave phase, not only does a magnetic impurity induce subgap YSR states, but also a potential impurity induces two-fold degenerate subgap bound states, as clearly illustrated in Fig.~\ref{M1.fig}(b). 
Intriguingly, potential impurities lead to the formation of bound states, although no level crossing takes place. 
We can understand the existence of sub gap states from potential impurities, using simple topological arguments. The potential impurity site represents a small, one site large, domain where we can define a local effective chemical potential $\mu_{\rm eff} = \mu + U$. Although the chemical potential is a formally global property, this local value is still very helpful to understand the physics of the system. The black dashed line in Fig.~\ref{M1.fig}(b) marks the value where $\mu_{\rm eff}$ is such that the impurity site, if considered by itself, enters the trivial phase. Clearly, this remarkably coincides with the impurity states starting to appear in the subgap region. Thus, the subgap states of a potential impurity are the edge states between a trivial region, defined by the impurity, and the surrounding topologically nontrivial bulk. The fact that the edge states are not at zero energy but found at higher energies, is attributed to a finite size effect, causing severe hybridization of the would-be edge states for single impurities. The results in the next subsection, directly connect these results to those of larger impurity domains.
As is clearly seen in the inset of Fig. \ref{M1.fig}(b), where we plot the subgap state as function of $1/U$ instead,  the subgap states energies not only avoid crossing at zero energy, even slightly tending to the edge of the superconducting gap when the scattering potential becomes very large, corresponding to a vacancy. The lack of level crossing implies that the superconducting gap will survive even for finite concentration of potential scatterers and in the thermodynamic limit, although the gap will be attenuated compared to a clean system.
It is possible to also connect this result with the Anderson's theorem, which states that any scatterings that preserve the time-reversal symmetry cannot break the Cooper pairing in $s$-wave SCs with fully isotropic gap functions in momentum space. Subsequently, bound states with an energy inside the gap cannot be induced due to the potential impurities in such conventional SCs. For the topological $s_\pm$-wave phase, a formal violation of the theorem is allowed since the order parameter is not fully isotropic. Our results shows that such violation also indeed take place, but at the same time, since the subgap states never cross zero energy for any value of the potential scattering strength, the Anderson's theorem is not strongly violated. Therefore superconductivity is also preserved in a dirty sample of a TRI topological SC, but with reduced gap and weakened superconducting correlations.

\begin{figure}
\includegraphics[width=0.23\textwidth]{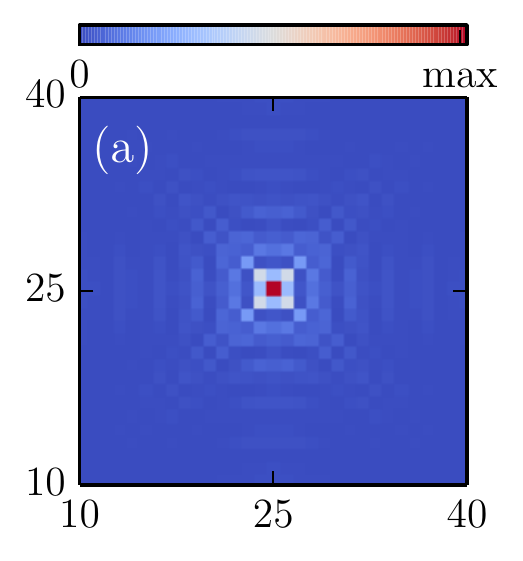}
\includegraphics[width=0.23\textwidth]{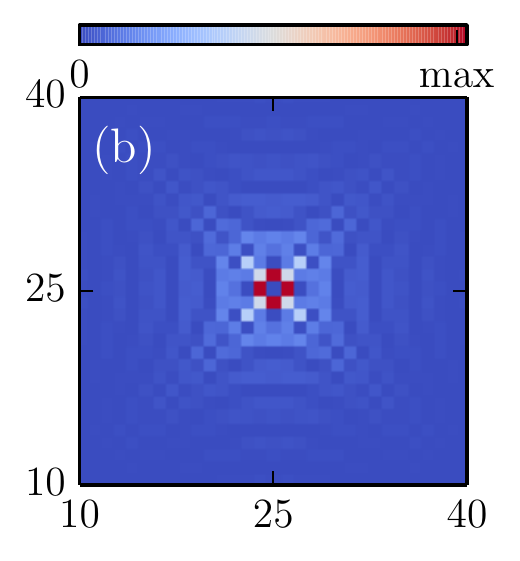}
\includegraphics[width=0.23\textwidth]{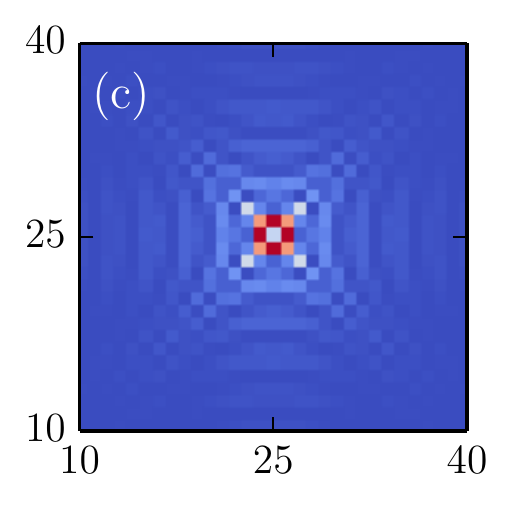}
\includegraphics[width=0.23\textwidth]{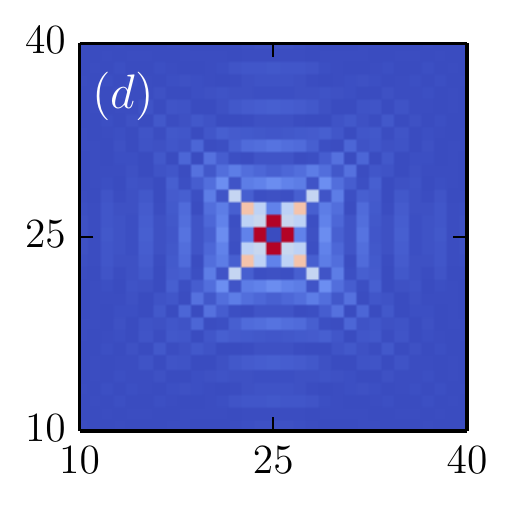}
\caption{(Color online.) Magnitude squared of electronic wave function of subgap bound states due to a potential impurity. $U =3$ potential impurity at negative (a) and positive (b) energy, as well as vacancy at negative (c) and positive (d) energy.}
\label{WaveFunction.fig}
\end{figure}

The subgap state from potential impurities in the topological non-trivial phase can be experimentally probed by using STM/STS. These techniques can not only confirm the appearance of subgap states, but can also scan the spatial distribution for both positive and negative biases. Previously, these kind of measurements for high-$T_c$ SCs has been performed for both potential \cite{Hudson2000} and magnetic \cite{Hudson2001} impurities and in both cases different spatial patterns were observed for positive and  negative biases.  The spatial extent of the subgap impurity states at both negative and positive bound states energies for the potential impurity is illustrated in Fig.~\ref{WaveFunction.fig}(a) and (b), respectively. These figures illustrate very strong localization of the bound states wave function to the impurity, although the tail of the bound state wave function extends along $y{=\pm}x$.
Moreover, the figures show an asymmetric behavior. 
While for the negative bias peak, as illustrated in  Fig.~\ref{WaveFunction.fig}(a), the maximum intensity occurs on top of the impurity itself and sharply decreases moving to its neighbors, for positive bias we see in Fig.~\ref{WaveFunction.fig}(b) how the intensity accumulates on the nearest neighbors to the impurity. This asymmetry is a result of the constant sum  of electron and hole parts in each BdG state. Thus, with the electronic DOS accumulated on top of the impurity for the negative energy bound state, the electronic DOS of the positive energy bound state appear instead on nearest neighbor sites. 
As we increase the potential scattering into the unitary limit, the impurity-induced states remain localized to the impurity site, as shown in Figs.~\ref{WaveFunction.fig}(c,d). This is notably in spite of the energy of the impurity state tending towards the gap edge, thus making the impurity-induced state not isolated in energy. In this limit there is naturally no LDOS on the impurity site, with the electron wave functions instead accumulate on the nearest neighbor sites.  
 
\subsubsection{Impurity island}
\begin{figure}[t]
\center
\includegraphics[width=0.4\textwidth]{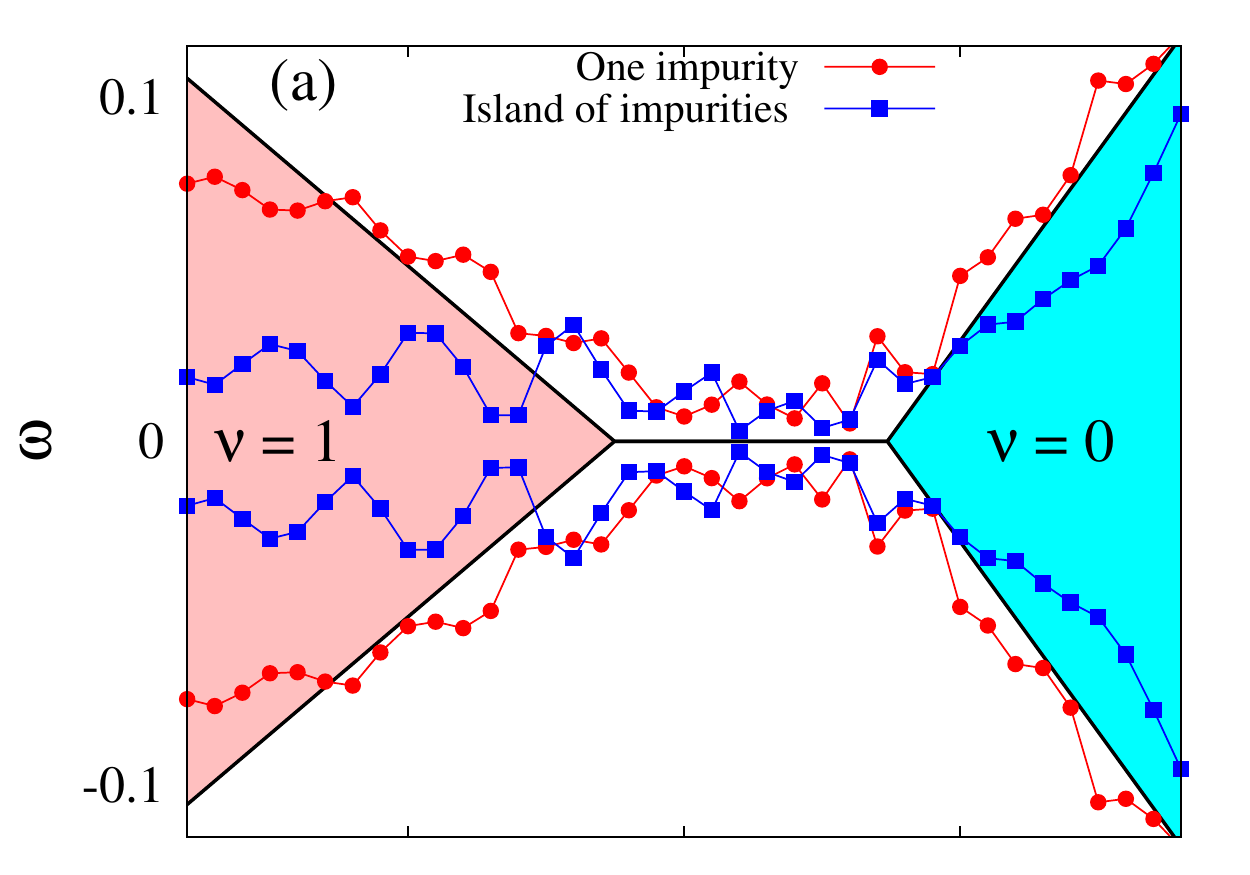}
\includegraphics[width=0.4\textwidth]{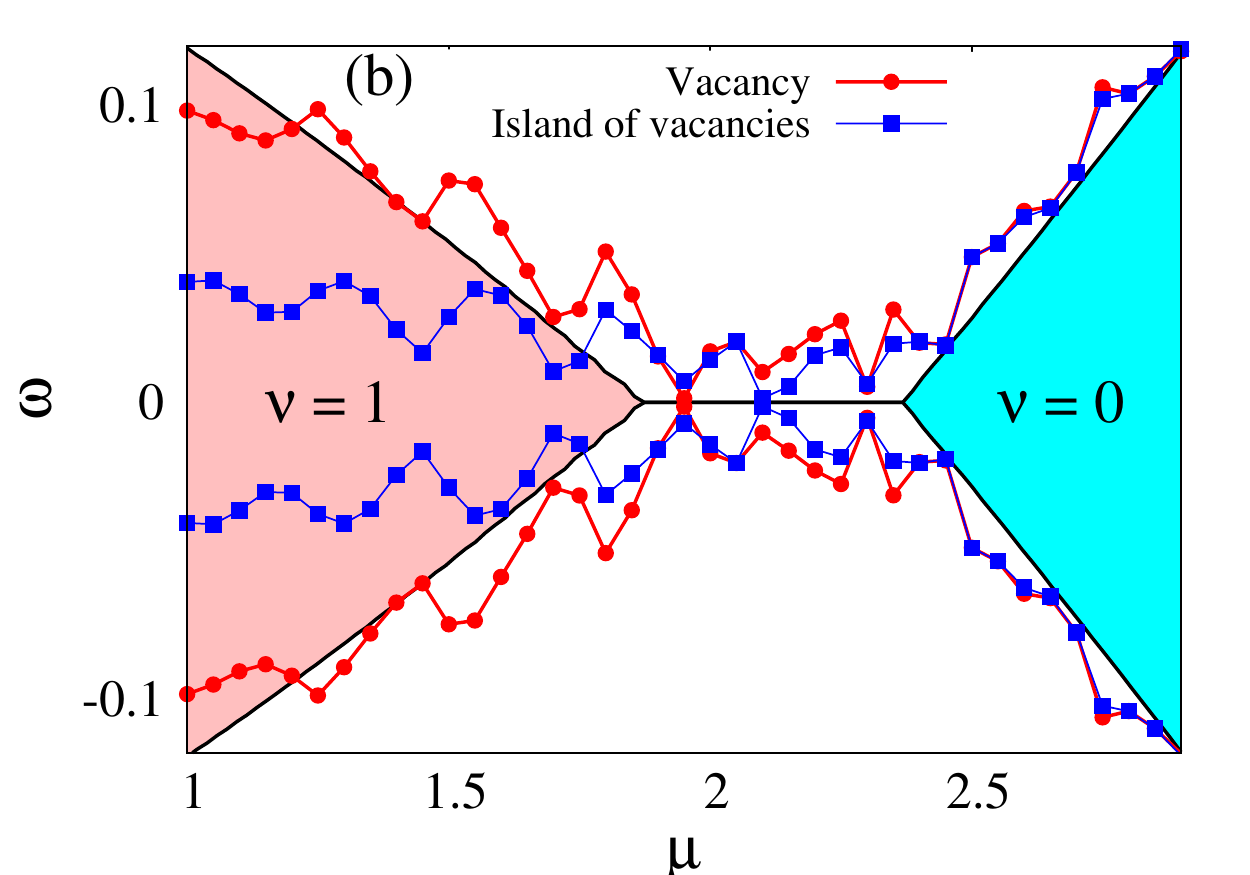}
\caption{(Color online.) Impurity bound states energy as a function of the chemical potential for (a) $U=4$ and (b) vacancy ($U\rightarrow \infty$) for a single impurity (red) and a $5 \time 5$ square island of impurities (blue). Tuning the chemical potential drive the impurity's host from topological non-trivial ($\nu=1$, pink) to trivial state ($\nu=0$, cyan) past the nodal region (white).}
\label{Fig7.fig}
\end{figure}
So far we have shown that a potential impurity only induces subgap bound states in the topological phase and can thus be used to distinguish between different topological classes in the under study system. However, these bound states do not go very deep inside the gap regardless of the impurity scattering strength.
The potential impurity in its extreme limit, i.e.~for a vacancy, can be seen as creating a topologically trivial area inside the host. In this case, the impurity subgap bound states can also be viewed as an edge effect between two different topological regions, non-trivial in the host, trivial on the vacancy. However, since the size of a single impurity is only one lattice site, strong finite quantization along the edge leads to states far from zero, albeit in the subgap regime. Here we investigate connecting the single impurity limit to that of proper edge states for a large hole inside the host material. We do this by increasing the spatial size of the impurity, where we thus expect the impurity bound states to appear deeper inside the gap since finite quantization effects decreases with increasing hole size \cite{Menard2017,Kris-thesis}.

In Fig.~\ref{Fig7.fig}(a) we compare the results for a single potential impurity of strength $U =4$ and an island of the same impurities as a function of chemical potential, which probes the topologically non-trivial phase to the topologically trivial phase. Since the potential scattering $U$ enters into the Hamiltonian as an additional local chemical potential, this choice of $U$ brings the impurity island locally into the trivial phase.
As seen, for a single potential impurity the subgap bound states in the non-trivial region appear relatively close to the band gap edge. However, for a square island composed of $5\times 5$ potential impurities the bond states are deep inside the gap. For this island, we only plot the lowest in energy bound states, but there are in fact more subgap states.
Concentrating on the trivial region, we can see that all the states closely follow the band gap, although for the impurity island, there are some states that are technically inside the gap. The deep subgap states for an impurity island in the topologically non-trivial phase are entirely dissimilar to their shallow counterpart in the trivial phase. 
\begin{figure}[bht]
\center
\includegraphics[width=0.23\textwidth]{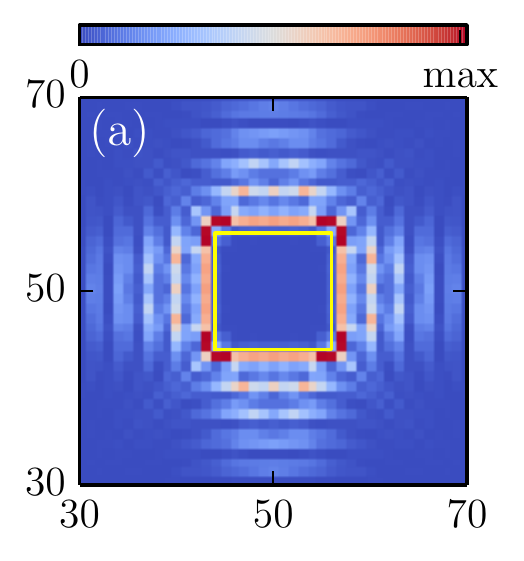}
\includegraphics[width=0.23\textwidth]{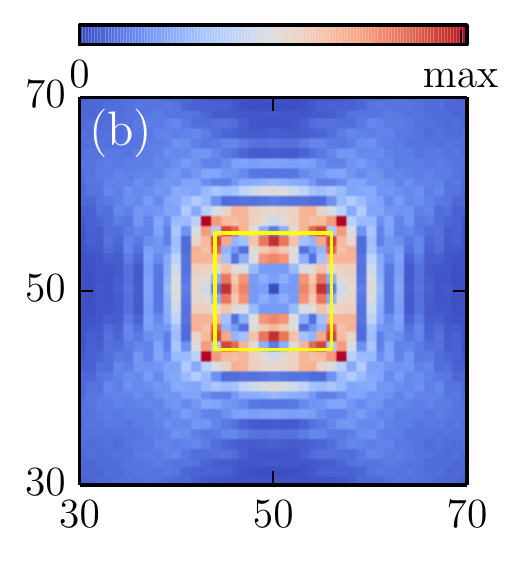}
\includegraphics[width=0.23\textwidth]{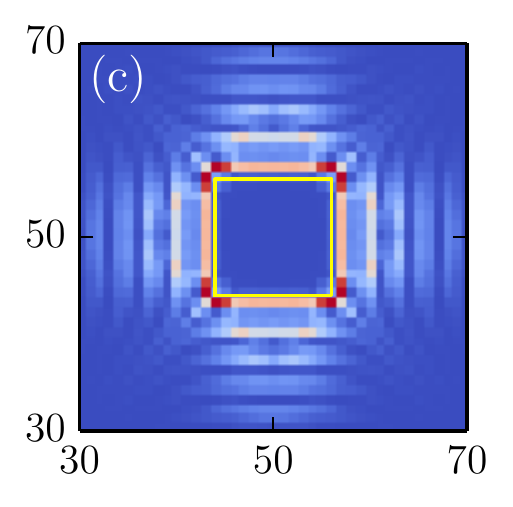}
\includegraphics[width=0.23\textwidth]{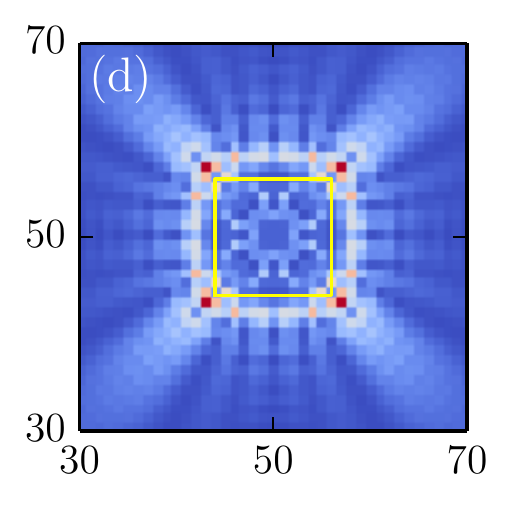}
\caption{(Color online.) Density (magnitude squared) of electronic wave function of subgap bound states due to a potential impurity island ($U =4$) for topologically non-trivial (a,c) and trivial (b,d) phases. Here $\mu=1.0$ and $2.7$ for topological and trivial phases respectively. (a,c) correspond to negative bias, (c,d) to positive bias, and the yellow square marks the boundary of the impurity island, with size $13 \times 13$ lattice points.}
\label{IslandWaveFunction.fig}
\end{figure}
 To shed more light on the difference of impurity scattering in topological and trivial phases, we plot the local density of states for subgap states in both phases.
As it is shown in the Fig.~\ref{IslandWaveFunction.fig}, while the subgap states in the topological phase
are strongly localized at the boundaries of the island (a,c), in trivial phase they are not localized at all and fully penetrate inside the island (b,d). So the states with energies slightly below the superconducting gap in the trivial phase are crucially different from the subgap states attributed to impurity effects. Putting this together with the fact that states with energies close to the gap edge can be veyr hard to distinguish in experiments, the strong subgap states in the  topological phase which are absent in the trivial phase can still be used as a true fingerprint of TRI topological SC phase.

Let us now return to  Fig.~\ref{Fig7.fig}(b) where we consider the unitary limit of scattering, i.e.~vacancies. In this limit, there is no bound state in the trivial region while there are deep subgap states for topological case, specially for the vacancy island which is then a small hole in the material. For even larger islands the lowest energy state will move even further down in energy, until it reaches zero. At the same time there will be a proliferation of subgap states, as there should be two propagating one-dimensional edge states with minimal finite size quantization surrounding a large hole. Note that the oscillatory behavior  of the impurities states, in Fig.~\ref{Fig7.fig}(a,b), is related to the finite size of the full lattice, i.e.~$L_x=L_y=51$ lattice points in this case. We expect a smoother behavior in the limit of $L\rightarrow \infty$.
We can therefore conclude that an island of potential impurities further enhances the dicotomy between topological and trivial phases as it deepens the energy of the subgap impurity bound states.

\section{Concluding remarks}\label{sec-conc}
In summary, we investigate the effect of magnetic and potential impurities on the electronic properties of a hybrid structure constructed from a Rashba layer in proximity to an extended $s$-wave superconductor. This time-reversal invariant system can undergo a topological phase transition from a trivial $s$-wave superconductor to a non-trivial phase with $s_\pm$-wave symmetry through a nodal phase. We find that a single magnetic impurity induces bound subgap states in both the topologically trivial and non-trivial phases. In contrast, a single potential impurity only induces bound subgap states in the topological phase. This marks a sharp difference between the topological $s_\pm$ SC and $s$-wave SCs in which non-magnetic impurities cannot induce any subgap state. In fact this offers explicit proof that potential impurities can produce in-gap states in $s$-wave SCs.
Interband scattering has previously been needed to explain why potential impurities generate subgap states in $s_\pm$-wave superconductors \cite{Matsumoto2009}. However, in our system there is no interband scattering, even from impurities. Thus our results reveal an entirely other mechanism for generating subgap states from potential impurities in extended $s$-wave superconductors. The mechanism in this work is intimately tied to topology, with subgap states only appearing in the non-trivial topological phase. To further strengthen the connection to topology, and particularly the associated helical edge states, we also study small islands of potential impurities. The impurity bound states appear progressively deeper within the energy gap for increasing size of impurity islands, but due to finite-size quantum confinement effects they are generally found at non-zero energy for all small islands. Still, note that subgap states appear even for single impurity sites. This behavior should be contrasted with the trivial phase, where neither single potential impurities nor impurity islands create notable subgap states, even in the strong scattering limit.

These results suggest that impurity effects provide a very valuable probe to detect time-reversal invariant topological superconductivity in such systems. Opposite to the major current tools, which use the edge states and related effects at the boundaries to study topological features, our work belong to another category in which the information about the topology is gathered from inside bulk.
This we hope will also trigger further investigations of time-reversal invariant superconducting topological states and bulk impurity effects from both theoretical and experimental perspectives. 

\acknowledgements
MM would like to thank D.~Kuzmanovski and A.~Ramires for fruitful discussions. AGM acknowledges financial support from the Iran Science Elites Federation under Grant No.~11/66332.
MM, ABS, and FP acknowledge financial support from the Swedish Research Council (Vetenskapsr\aa det, Grant No.~621-2014-3721), the G\"{o}ran Gustafsson Foundation, the Swedish Foundation for Strategic Research (SSF), the Wallenberg Academy Fellows program, and the Knut and Alice Wallenberg Foundation.

\end{document}